\documentclass[twocolumn,prb,aps,superscriptaddress]{revtex4-2}
\usepackage{amsmath}   
\usepackage{amssymb}
\usepackage{amsfonts}
\usepackage{mathrsfs}
\usepackage{graphicx}
\usepackage{xcolor}

\usepackage{siunitx} % Required for alignment

\begin{document}

\title
{Piezoplasmonics: strain-induced tunability of plasmon resonance in AlAs quantum wells}

%\author{A.~R.~Khisameeva$^{a,b}$, V.~M.~Muravev$^{a}$, I.~V.~Kukushkin$^{a}$}
%\affiliation{$^a$Institute of Solid State Physics, RAS, Chernogolovka, 142432 Russia \\
%$^b$ Moscow Institute of Physics and Technology, Dolgoprudny, 141700 Russia}
\author{A.~R.~Khisameeva}
\affiliation{Moscow Institute of Physics and Technology, Dolgoprudny, 141700 Russia}%
\affiliation{Institute of Solid State Physics, RAS, Chernogolovka, 142432 Russia}%

\author{V.~M.~Muravev}%
\author{I.~V.~Kukushkin}%

\email[Corresponding e-mail:]{ muravev@issp.ac.ru}

\affiliation{Institute of Solid State Physics, RAS, Chernogolovka, 142432 Russia}%

\date{\today}

\begin{abstract}
We demonstrate tuning of two-dimensional (2D) plasmon spectrum in modulation-doped AlAs quantum wells via the application of in-plane uniaxial strain. We show that dramatic change in the plasma spectrum is caused by strain-induced redistribution of charge carriers between anisotropic $X_x$ and $X_y$ valleys. Discovered piezoplasmonic effect provides a tool to study the band structure of 2D systems. We use piezoplasmonic effect to measure how the inter-valley energy splitting depends on the deformation. This dependency yields the AlAs deformation potential of $E_2 = (5.6 \pm 0.3)$~eV.
\end{abstract}

\maketitle

In the past few decades, semiconductor electronics has gone through an impressive development. However, further progress in the working frequency of the devices has been hampered by fundamental physical limitations. The transistor cut-off frequency can be estimated as $f_c \sim v_s/2 \pi L$, where $v_s$ is a saturated drift velocity and $L$ is the transistor gate length~\cite{Pengelly:94}. For a typical industrial GaAs HEMT having a gate length of about $0.1$~$\mu m$ the frequency is restricted by $f_c \approx 100$~GHz. Thus, there is an active search for novel high-frequency device concepts. One of the possibilities is the use of two-dimensional (2D) plasma waves as carriers of electric signals. Indeed, the velocity of plasma excitations in two-dimensional electron systems can reach $v_p = 10^9$~cm/s, which is two orders of magnitude higher than the maximally attainable electron-drift velocity. Therefore, plasmon frequency can reach the terahertz (THz) range for gate lengths of a micron size. It is this latter capability that has generated interest in 2D plasmonic devices such as THz detectors~\cite{Peralta:02, Knap:02, Shaner:05, Muravev:12}, emitters~\cite{Tsui:80, Dyakonov:93, Mikhailov:98, Knap:04, Otsuji:08}, and field enhancement structures~\cite{Koppens:18, Lee:19, Smith:19}. 

The frequency of 2D plasmons follows the formula~\cite{Stern:67}: 
 \begin{equation}
\omega_p (q)=\sqrt{\frac{n_s e^2 q}{2m^{\ast} \kappa_0  \kappa (q)}}  \qquad (q \gg \omega/c).
\label{2D}
\end{equation}
Here, $q$ is the wave-vector of the plasmon, while $n_s$ and $m^{\ast}$ are the density and the effective mass of the 2D electrons. The vacuum permittivity and the effective permittivity of the surrounding medium are denoted as $\kappa_0$ and $\kappa (q)$, respectively. One of the most attractive properties of 2D plasmons is that their frequency can be tuned in wide range by changing the electron density. The tunability of 2D plasmons is vital for the field of frequency-agile integrated circuits based on plasmonics. 

\begin{figure*}[t!]
%\PSbox{Pict2.eps}{8.5cm}{7.7cm}
\includegraphics[width=0.8 \textwidth]{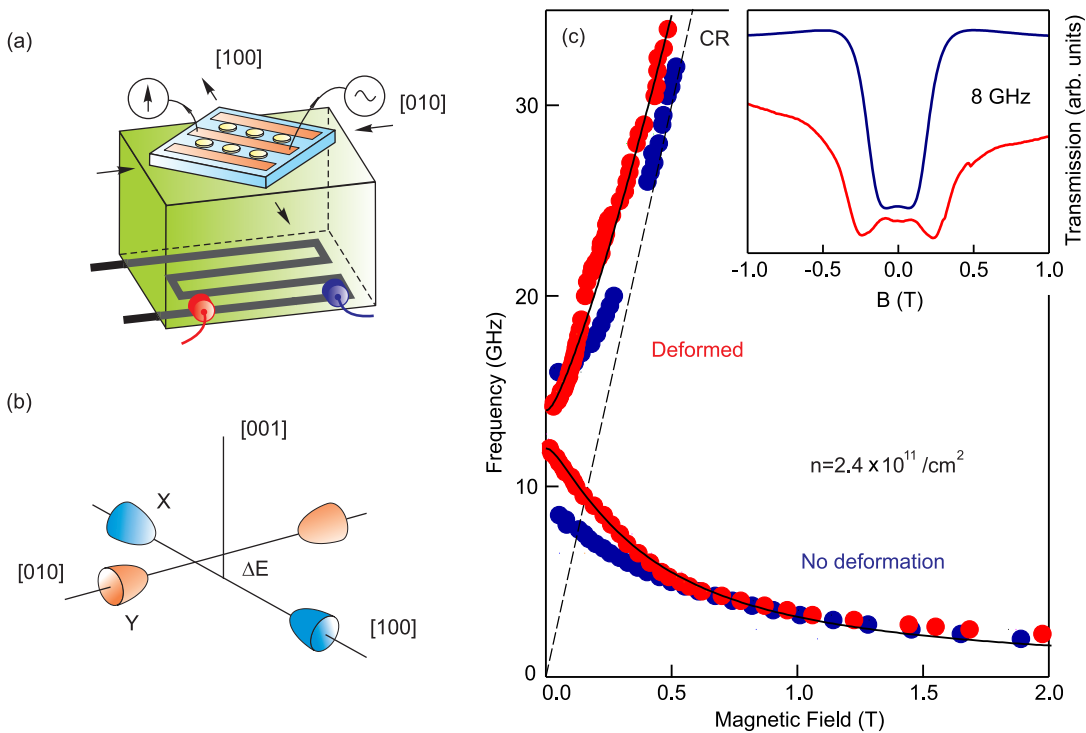}
\caption{ (a) Schematics of the experimental set-up. (b) Illustrative drawing of the AlAs Fermi surface in the plane of the quantum well. (c) Magnetodispersion of the two-dimensional plasma excitations in AlAs disks for the non-deformed sample (blue dots), and the same sample glued on the piezo-stack (red dots). In the latter case, the sample is already strained at zero piezo bias because of the piezo-stack compression during the cooling. Dashed line represents the cyclotron frequency $\omega_c = eB/\sqrt{m_{\rm l} m_{\rm tr}}$, and solid lines indicate magnetodispersion dependences according to Eq.~(\ref{SM_1}). Inset shows coplanar waveguide transmission at $f=8$~GHz for these two cases. The total 2D density is kept fixed $n_s=2{.}4\times 10^{11}\,\text{cm}^{-2}$.}
\label{Fig1}
\end{figure*}

AlAs two-dimensional electron system (2DES) has promising prospective for research and practical realization of plasmonic concepts~\cite{Shayegan:06}. This composite is an indirect band-gap semiconductor with conduction band valleys centered at the X-points of the Brillouin zone, along [100], [010], and [001] crystallographic directions. Hence, they are traditionally referred to as $X_x$, $X_y$, and $X_z$ valleys, respectively. The corresponding anisotropic, elliptical Fermi contours are characterized by a heavy longitudinal mass $m_{\rm l} = 1.1 m_0$ and a light transverse mass $m_{\rm tr}= 0.2 m_0$, where $m_0$ is the free electron mass. Because of the GaAs–AlAs lattice mismatch, only the two valleys $X_x$ and $X_y$ lying in the plane of the 2DES are occupied in AlAs quantum wells wider than $6$~nm. Moreover, the residual in-plane strain lifts the $X_x$ and $X_y$ valley degeneracy, leading to inter-valley energy splitting $\Delta E$ (see Fig.~\ref{Fig1}(b)). The collective plasma excitations in the 2DES with anisotropic energy spectrum possess a number of unique properties. In particular, the discovery of a gap in the spectrum of plasma excitations in perfectly symmetric circular samples~\cite{Muravev:15, Khisameeva:18, Khisameeva:20}. It was shown that inter-valley energy splitting significantly influences the plasmon spectrum through redistribution of charge carriers between $X_x$ and $X_y$ valleys. However, in all aforementioned experiments, the built-in uniaxial deformation has remained a hidden undefined parameter, partly due to the difficulty of measuring it, partly since the built-in deformation depends on lots of factors.

It is widely believed that strain has a negligible effect on the properties of plasmons in solids. Indeed, typical deformation that can be achieved by external means is on the order of $\varepsilon = 10^{-4}$, which translates into a variation in the 2D-density $\Delta n_s / n_s = 2 \times 10^{-4}$ with corresponding negligible change in the plasmon frequency $\Delta f_p/f_p = 10^{-4}$. However, there is another way the strain can affect the plasmon spectrum –-- when deformation modifies the electronic structure of a solid. In this paper, we report on measurements of the plasmon spectrum in AlAs quantum wells as a function of controlled uniaxial deformation. Our findings demonstrate that, at low temperatures, the 2D plasmon spectrum can be tuned over a wide range by means of externally applied stress. The experimental data can be explained quantitatively by a simple model that takes into account the stress-induced transfer of charges between the valleys. The developed device has been used to measure the AlAs deformation potential.

The experiments were carried out on high-quality AlAs/Al$_{x}$Ga$_{1-x}$As ($x=0.46$) heterostructure containing quantum well of width $W = 15$~nm. The quantum well was grown along the $[001]$ direction on undoped GaAs via molecular beam epitaxy. The two-dimensional electron density $n_s$ and the low-temperature transport mobility $\mu$ were $2{.}4\times 10^{11}\,\text{cm}^{-2}$ and $1{.}5\times 10^{5}\,\text{cm}^{2}$/Vs, respectively. A coplanar waveguide (CPW) was fabricated on the sample surface using the photolithography method. Three equidistant disks hosting 2DES were made at each slot of the CPW. The diameter of the disks was $d=0.5$~mm with a distance between them of $1.5$~mm (Fig.~\ref{Fig1}(a)). The CPW was oriented along $[110]$ crystallographic direction. The total CPW length was $4$~mm, with the width of the central strip of $1.1$~mm and slots of $0.6$~mm. The experimental method relied on the absorption of microwave radiation passing through the coplanar waveguide~\cite{Andreev:12, Muravev:15}. The resonant absorption of the microwaves occurs whenever a plasmon is excited in the disks. Microwave radiation ($f = 1-40$~GHz) was guided to the input port of the CPW via 50~$\Omega$ coaxial cable. The second coaxial cable delivered the radiation from the output of the CPW to the Herotek DTA1-1880A tunnel diode. The diode was placed outside the cryostat. To apply tunable strain we glued the sample with CPW to one side of a piezo-stack actuator (PSA)~\cite{Shayegan:03, Shayegan:04, Shayegan:04-2, Grayson:11, Grayson:12}. The $[010]$ crystallographic direction was aligned along the stroke direction of the actuator. We  thinned the sample up to $\approx 0.2$~mm to fully transmit deformation from PSA to the sample. We used a resistance gauge to measure the applied strain (we refer to Supplemental Material for more details). The gauge was glued on the opposite side of the actuator. The sample was compressed (expanded) along the stroke direction by applying a bias voltage from $V=-200$ to $400$~V. All our experiments were carried out at a base sample temperature of $1.5$~K. 

%AlAs is an indirect gap semiconductor with conduction band valleys centered at the X-points of the Brillouin zone along [100], [010], and [001] crystallographic directions. These valleys are traditionally referred to as $X_x$, $X_y$, and $X_z$, respectively~\cite{Shayegan:06}. The corresponding anisotropic, elliptical Fermi contours are characterized by a heavy longitudinal mass $m_{\rm l} = 1.1 m_0$ and a light transverse mass $m_{\rm tr}= 0.2 m_0$, where $m_0$ is the free electron mass. Because of the GaAs–AlAs lattice mismatch, only the two valleys $X_x$ and $X_y$ lying in the plane of the 2DES are occupied in AlAs quantum wells wider than $6$~nm. Moreover, the residual in-plane strain lifts the $X_x$ and $X_y$ valley degeneracy, leading to inter-valley energy splitting $\Delta E$ (see Fig.~\ref{Fig1}(b)). Uniaxial external stress can tune energy splitting $\Delta E$ and thereby causes a redistribution of electrons between the valleys~\cite{Shayegan:03, Shayegan:04, Shayegan:04-2, Grayson:11, Grayson:12}.

If both $X_x$ and $X_y$ valleys are occupied, then dispersion of plasmons in the disk-shaped 2DES can be considered using a two-component anisotropic plasma model~\cite{Chaplik:81}. In this case, plasma oscillation in the disk can be factorized into two normal modes with frequencies:

\begin{equation}     
\Omega_{[100]}^2=\frac{e^2 q}{2 \kappa_0 \kappa^{\ast}} \left( \frac{n_{\rm x}}{m_{\rm l}}+\frac{n_{\rm y}}{m_{\rm tr}} \right), 
\label{3}
\end{equation} 

\begin{equation}     
\Omega_{[010]}^2=\frac{e^2 q}{2 \kappa_0 \kappa^{\ast}} \left( \frac{n_{\rm x}}{m_{\rm tr}}+\frac{n_{\rm y}}{m_{\rm l}} \right). 
\label{4}
\end{equation}
Here $\kappa^{\ast} = (\kappa_{\rm GaAs} + 1)/2$ is the effective dielectric permittivity of the surrounding medium and $q = 2.4/d$ is the wave vector in the disk geometry~\cite{Kukushkin:03}. Using these expressions, one can see that maximum gap in the plasmon spectrum occurs when one of the valleys $X_x$ or $X_y$ is empty. Whilst the gap collapses when the valleys are filled equally ($n_{\rm{x}}=n_{\rm{y}}$). If the total electron density $n_s = n_{\rm{x}} + n_{\rm{y}}$ in two valleys is fixed, then plasma frequencies $\Omega_{[100]}$ and $\Omega_{[010]}$ from Eqs.~(2-3) gives us density of electrons in each of the valleys $n_{\rm{x}}$ and $n_{\rm{y}}$. From the difference of these densities $\Delta n = n_{\rm{x}} - n_{\rm{y}}$, we can directly determine the inter-valley energy splitting $\Delta E$ (Fig.~\ref{Fig1}) using the 2D density of states~\cite{Muravev:15}:

\begin{equation}     
\Delta E=\frac{\pi \hbar^2 \Delta n}{\sqrt{m_{\rm l}m_{\rm tr}}}.  
\label{5}
\end{equation}

\begin{figure}[t!]
%\PSbox{Pict2.eps}{8.5cm}{7.7cm}
\includegraphics[width=0.47 \textwidth]{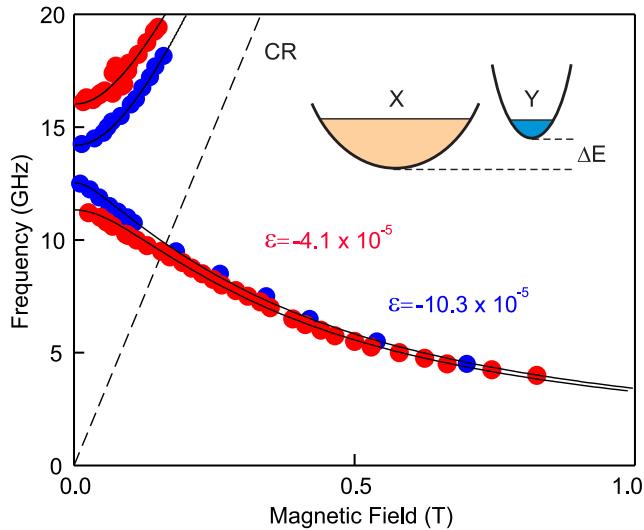}
\caption{Spectra of 2D magnetoplasma excitations in AlAs QWs measured for deformations of $\varepsilon_{[010]} = -10.3 \times 10^{-5}$ (blue dots) and $\varepsilon_{[010]} = -4.1 \times 10^{-5}$ (red dots). Inset displays a schematic drawing of the $X_x$ and $X_y$ valleys. Dashed line represents the cyclotron frequency $\omega_c = eB/\sqrt{m_{\rm l} m_{\rm tr}}$, and solid lines indicate magnetodispersion dependences according to Eq.~(\ref{SM_1}).}
\label{Fig2}
\end{figure}

The plasmon spectrum experiences a significant change under the action of uniaxial strain. Indeed, the inset to Fig.~\ref{Fig1}(c) shows magnetic-field dependencies of the coplanar waveguide transmission for the same sample without (upper curve) and with (bottom curve) deformation. In the latter case, the difference in thermal extension coefficients of semiconductor crystal and piezo actuator leads to the uniaxial compression of $\varepsilon_{[010]} = -8 \times 10^{-5}$ even at zero bias applied to the piezo actuator. Both traces are recorded at the same frequency of $f=8$~GHz.

To illustrate the effect of deformation on the plasmon spectrum, inset to Fig.~\ref{Fig1}(c) shows the magnetic-field dependencies of the coplanar waveguide transmission at $f=8$~GHz for the native sample with no deformation (blue curve) and the same sample glued to the piezo actuator (red curve). Each curve shows a resonance corresponding to the excitation of a standing plasma wave in the disk. We used a B-field spectroscopy to determine plasmon frequencies in the 2DES due to fact that this approach eliminates the problem of matching the microwave tract. The procedure for analyzing B-field transmission is detailed in the Supplemental Material II. The resulting magnetodispersion of plasma waves excited in the disks is shown in Fig.~\ref{Fig1}(c). Each magnetodispersion has two branches separated by a frequency gap at zero magnetic field. The gap at $B=0$ vividly manifests the highly anisotropic nature of the Fermi surface in AlAs 2DES. The low-frequency branch of the magnetodispersion in Fig.~\ref{Fig1}(c) corresponds to the excitation of edge magnetoplasmon (EMP)~\cite{Volkov:88}. Qualitatively this mode originates from skipping-orbit collective motion of electrons along the edge of the 2DES. The EMP frequency decreases as $\omega_{-} \approx \sigma_{\rm xy} q \propto n_{\rm s} q/B$ regardless of the zone structure. The fact that both EMP branches (red and blue dots in Fig.~\ref{Fig1}(c)) merge into a single asymptotic for $B>0.5$~T indicates that deformation does not affect the 2DES density. The high-frequency branch in Fig.~\ref{Fig1}(c) corresponds to excitation of the cyclotron magnetoplasma mode. Qualitatively this mode arises from collective motion of electrons in cyclotron orbits throughout the entire area of the 2DES. It tends to the cyclotron resonance line $\omega_{\rm c} = e B/\sqrt{m_{\rm l} m_{\rm tr}}$ in the limit of strong magnetic field (dashed line marked by CR in Fig.~\ref{Fig1}(c)). Magnetic-field behavior of both edge and cyclotron modes can be analytically described using the dipole approximation (for details see Supplemental Material III)~\cite{Allen:83, Dahl:91, Shikin:91, Margulis:01}
\begin{multline}
\omega_{\pm}=\frac{1}{2} \Big[ \sqrt{(\Omega_{[100]}+\Omega_{[010]})^2+\omega_c^2} \\
\pm \sqrt{(\Omega_{[100]}-\Omega_{[010]})^2+\omega_c^2} \Big],
\label{SM_1}
\end{multline}     
Black lines in Fig.~\ref{Fig1}(c) represent fitting of Eq.~(\ref{SM_1}) to experimental data. It is obvious that approach works well and provides a tool for determining plasma frequencies at zero magnetic field.

\begin{figure}[t!]
%\PSbox{Pict2.eps}{8.5cm}{7.7cm}
\includegraphics[width=0.47 \textwidth]{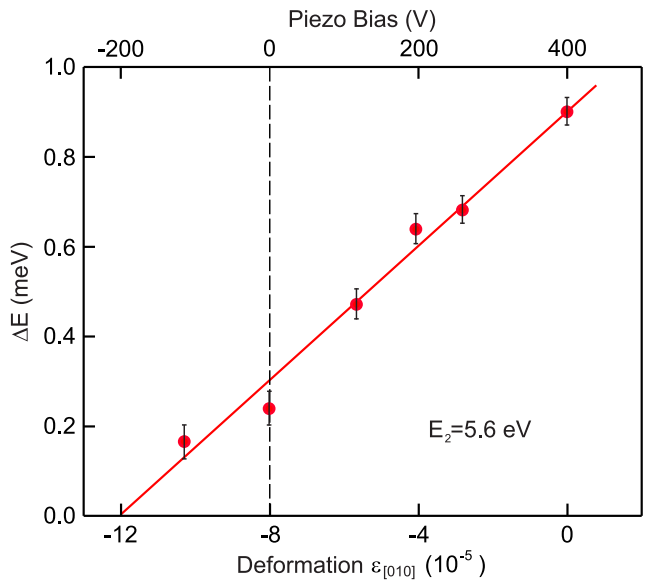}
\caption{Inter-valley energy splitting $\Delta E$ between in-plane $X_x$ and $X_y$ valleys as a function of uniaxial deformation $\varepsilon_{[010]}$. A linear fit shown by the solid line yields a shear deformation potential of $E_2 = 5.6$~eV. Experimental points are shown with error bars. The dashed line marks the compression generated by the difference in thermal extension coefficients between the semiconductor crystal and piezo-stack.}
\label{Fig3}
\end{figure}  

The magnetodispersion curve for unstrained sample (see Fig. \ref{Fig1}(c), blue dots) was studied in our previous work \cite{Muravev:15}, where the following zero-field frequencies were extracted: $\Omega_{[100]}/2 \pi = (8.9 \pm 0.2)$~GHz and $\Omega_{\rm{[010]}}/2 \pi = (15.7\pm0.2)$~GHz, and $\Delta E = (0.90 \pm 0.05)$~meV. In the present paper we apply the same procedure for a strained sample (red dots in Fig.~\ref{Fig1}(c)), then we arrive at the following value of the inter-valley splitting $\Delta E (\varepsilon_{[010]} = -8 \times 10^{-5}) = (0.14 \pm 0.01)$~meV. Taken together, these results demonstrate that uniaxial strain leads to a significant change in inter-valley splitting $\Delta E$, which translates into a dramatic modification of the 2D plasma spectrum.

One of the most outstanding properties of 2D plasmons is  that  their  frequency  can  be  tuned. Most often, the tunability of  2D  plasmons is implemented via change of electron density by a voltage on a gate electrode. Discovered piezoplasmonic effect opens up an additional way for tuning of 2D plasmons. To demonstrate this, we performed a set of experiments with different voltage $V$ applied to the piezo actuator. Fig.~\ref{Fig2} shows 2D plasmon magnetodispersion for two cases: when $V=-110$~V (blue dots), and $V=200$~V (red dots). Corresponding values of deformation values are $\varepsilon_{[010]} = -10.3 \times 10^{-5}$ and $\varepsilon_{[010]} = -4.1 \times 10^{-5}$.

Found piezoplasmonic effect is a powerful tool to study the band structure of semiconductors. Indeed, based on Eqs.~(\ref{3} - \ref{5}) one can calculate inter-valley splitting $\Delta E$ for each value of deformation. For example, for curves in Fig.~\ref{Fig2} the corresponding values of inter-valley splitting are $\Delta E (\varepsilon_{[010]} = -10.3 \times 10^{-5}) = 0.17$~meV and $\Delta E (\varepsilon_{[010]} = -4.1 \times 10^{-5}) = 0.64$~meV. The resulting dependency of $\Delta E$ on applied deformation $\varepsilon_{[010]}$ is shown in Fig.~\ref{Fig3}. The energy spacing between $X_x$ and $X_y$ valleys changes linearly with deformation. From this linear dependency, we can deduce the shear deformation potential of AlAs. Indeed, the shear strain $\epsilon$ in our experiment equals to the difference of deformations along the [010] and [100] directions: $\epsilon = \varepsilon_{[010]} - \varepsilon_{[100]}$. Since the Poisson ratio of AlAs is $r=0.32$~\cite{Shayegan:03}, then $\epsilon = (1+r) \varepsilon_{[010]}$. The inter-valley energy splitting is then $\Delta E = E_2 \times (1+r) \varepsilon_{[010]}$. Thus, according to the experimental data in Fig.~\ref{Fig3}, we arrive at the following value of the shear deformation potential $E_2 = (5.6 \pm 0.3)$~eV. This value is in good agreement with previous studies~\cite{Kettles:91, Shayegan:06}.

%Finally, we note that at the point where both valleys align, we still observe a small but appreciable gap in the electron spectrum $\Delta E = E_{\rm SAS}= (0.12 \pm 0.03)$~meV. The physical origin of this gap is not clear and needs further research. However, one of the most probable explanations is quantum interference between wave functions of electrons in $X_x$ and $X_y$ valleys. As a result, symmetric and antisymmetric states arise from the ground states of the two valleys. These states are separated by the energy gap $E_{\rm SAS}$~\cite{???}.   

%Figure~\ref{Fig3} presents a resultant experimental dependency of inter-valley energy splitting $\Delta E$ on applied external uniaxial deformation $\varepsilon_{[010]}$. The dashed line marks the value of residual deformation due to the difference in thermal extension coefficients at $T= 4.2$~K. The value of valley splitting $\Delta E = (0.90 \pm 0.05)$~meV is corresponds to deformation $\varepsilon_{[100]} = 0$ when the sample was not glued on PSA. Our results demonstrate the increase of inter-valley splitting by almost an order of magnitude in controllable way, that makes such 2DES a versatile subject for valleytronics. 

In summary, we have investigated plasma and magnetoplasma excitations in 2DES based on AlAs QWs under uniaxial strain provided by the piezoelectric actuator. The external strain has changed in a controllable way the energy splitting between $X_x$ and $X_y$ valleys. We have shown that redistribution of charge carriers between the valleys has led to a dramatic modification of the 2D plasma spectrum. The discovered piezoplasmonic effect proved to be a useful probe of the electronic properties of the 2DES. We used it to establish how the inter-valley energy splitting depends on the uniaxial deformation.  

\subsection{Supplementary Material}

Supplementary Material I includes details of the piezo-stack calibration procedure. Supplementary Material II provides more experimental data and analysis of the resonance line-shape and position. Supplementary Material III presents the basics for calculation of 2D magnetoplasmon dispersion in the case of the anisotropic energy spectrum.

\subsection{Acknowledgments}

The authors gratefully acknowledge the financial support from the Russian Science Foundation Grant No.~19-72-30003 (development of the experimental technique) and Grant No.~18-72-10072 (analysis of the experimental data). We would also like to thank C.~Reichl, W.~Dietsche, and W.~Wegscheider for the  provided AlAs structures.

The data that support the findings of this study are available from the corresponding author upon reasonable request.

\end{document}